\documentclass{symp240}
\usepackage{graphicx}

\title[RZ~Cassiopeia: an Eclipsing Binary with a Pulsating Component] 
{RZ~Cassiopeia: an Eclipsing Binary with a Pulsating Component}

\author[A. Golovin \& E. Pavlenko]   
{Alex Golovin$^1$%
  \thanks{Present address: Kyiv National Taras Shevchenko University, Volodymyrska Street, 64, Kyiv 01033, Ukraine},
\and Elena P. Pavlenko$^2$}

\affiliation{$^1$ Main Astronomical Observatory of NAS of Ukraine;
Zabolotnoho str.,\\ 27, Kyiv, 03680, Ukraine \break Kyiv National
Taras Shevchenko University, Volodymyrska Street, 64, Kyiv, 01033,
Ukraine; \break Visiting astronomer of the Crimean Astrophysical
Observatory, Nauchny, Crimea, Ukraine
\break email: astronom\_ 2003@mail.ru\\[\affilskip]
$^2$Crimean Astrophysical Observatory, p/o Nauchny 19/17, Crimea,
98409, Ukraine}

\pubyear{2007}
\volume{240} 
\pagerange{119--126}
\date{?? and in revised form ??}
\jname{Binary Stars as Critical Tools \& Tests\\ in Contemporary Astrophysics}
\editors{W.I.\ Hartkopf, E.F.\ Guinan \& P.\ Harmanec, eds.}

\begin{document}

\maketitle

\begin{abstract}
We report time-resolved $V$- and $R$-band CCD photometry of the eclipsing
binary \object{RZ~Cas} obtained with 38-cm Cassegrain telescope at the
Crimean Astrophysical Observatory during July 2004 -- October 2005.
The obtained lightcurves clearly demonstrates rapid pulsations with
a period of about 22 min. A periodogram analysis of these oscillations
is also reported. On 12 January, 2005 (JD 2453383) we
observed rapid variability with higher amplitude ($\sim$0$^{\rm m}$\llap.1), 
that perhaps may be interpreted as a high-mass-transfer-rate event
and inhomogeneity of the accretion stream.       Follow-up
observations (both photometric and spectroscopic) of RZ~Cas are
strictly desirable for more detailed study of such events.
\end{abstract}

\firstsection 
\section{Introduction}

The A3V+K0 IV eclipsing binary RZ~Cas is an active semi-detached Algol system showing complex 
features in its lightcurve. It is well-known for demonstrating rapid pulsations (P $\sim 22$ min)
superimposed over its orbital eclipsing lightcurve. A brief overview of  previous investigations of 
RZ~Cas can be found in Golovin \& Pavlenko (2006).

\section{Observations}
RZ~Cas was observed in July 2004 -- October 2005 in the $V$ and $R$ bands from the Crimean 
Astrophysical Observatory (Ukraine) by the authors, using a 38-cm Cassegrain telescope equipped with 
an SBIG ST-7 CCD camera, cooled by a Peltier system to about $-20^{\circ} C$. The field-of-view 
covered a sky region of $12'\times8'$ and the pixel size was 0$''$\llap.9 $\times$ 0$''$\llap.9. The 
exposure time was 2$^{\rm s}$\llap.5  for the $R$ band and 5 or $10^{\rm s}$ for the $V$ band. To 
minimize dead time we used binning by a factor of 2. Data reduction was done using the ``Maxim DL" 
package. Reduction included bias, dark-frame subtraction and flat fielding using twilight sky 
exposures. Since the field of RZ~Cas is not crowded, the technique of aperture photometry was 
applied to extract the differential magnitudes. The total number of useful frames was 4365. The 
brightness of RZ~Cas was measured with respect to \object{GSC 4317$-$1578} ($\alpha = 02
^{\circ}48'41''.58;~ \delta = +69^{\circ}35'31''.3; J2000.0)$, while \object{GSC 4317$-$1437} 
($\alpha = 02^{\circ}48'38''.29;~ \delta = +69^{\circ}37'29''.9; J2000.0)$ served as a check star.
Unfortunately, no suitable comparison and check star could be including in the frames (stars are at 
least $4^{\rm m}$ fainter than the variable star).  The photometric error (determined from the 
difference \emph{check star -- comparison star}) is about 0$^{\rm m}$\llap.01. Figure~1 illustrates 
a 15$'$ $\times$ 15$'$ image of the RZ~Cas region from DSS. Variable, comparison and check stars are 
marked.

\begin{figure}
\begin{center}
\includegraphics[width=2.5in]{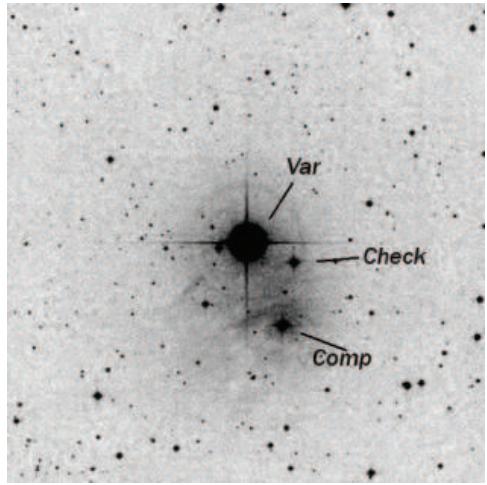}
  \caption{Finding chart.}
  \end{center}
\end{figure}

\begin{figure}[h!]
\begin{center}
\includegraphics[width=3in]{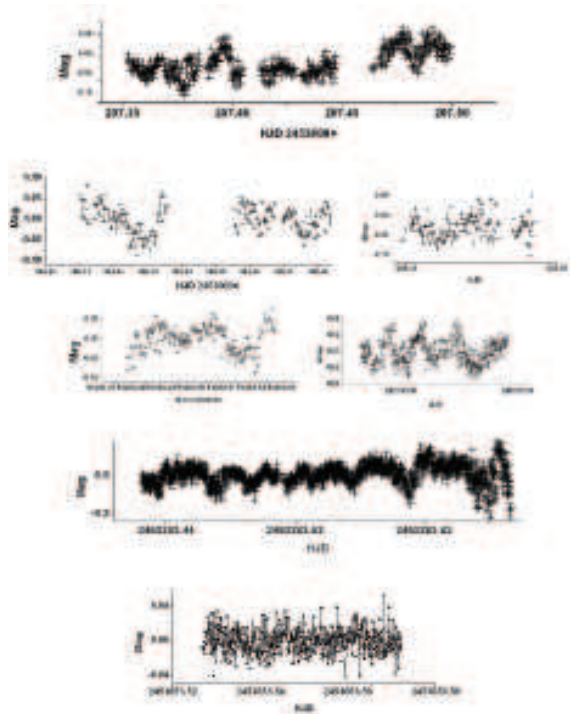}
\caption{Individual lightcurves.}
\end{center}
\end{figure} 

To rule out the possibility of observing variations caused by the comparison star, independent 
photometry of comparison star GSC 4317$-$1578 was performed with respect to check star GSC 4317$-$1437.

\section{Discussion} 
Figure~2 shows brightness variations during the maximal part of the lightcurve. Separate observing 
runs are manifested. As could be seen, pulsation amplitude could be assumed to be about
0$^{\rm m}$\llap.05 on average. A great deal of interest was caused by the fact that on 12
January, 2005 (JD 2453383) we observed rapid variability with higher amplitude 
($\sim$0$^{\rm m}$\llap.1), that perhaps may be interpreted as a high-mass-transfer-rate event and  
inhomogeneity of the accretion stream.   Periodogram analysis (Figure~3) reveals the following
periods: For 12, January, $2005: \\
P_1 = 0.041895 \pm 0.000849$   $(F_1 = 23.87 \pm 0.48)$ \\ $P_2 =
0.021238 \pm 0.000218$   $(F_2 = 47.08
\pm 0.48)$ \\ $P_3 = 0.017470 \pm 0.000148$  $(F_3 = 57.24 \pm 0.48)$ \\
$P_4 = 0.014323 \pm 0.000099$  $(F_4 = 69.82 \pm 0.48)$

\begin{figure}
\begin{center}
\includegraphics[width=3.5in]{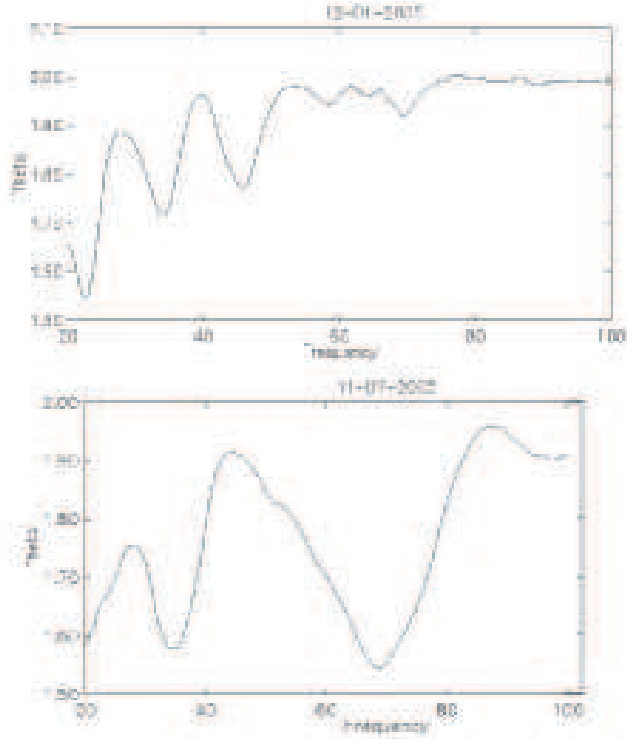}
  \caption{Periodograms.}
  \end{center}
\end{figure}

For all the other remaining nights: $P_1 = 0.014489 \pm 0.000169$
($F_1 = 69.02 \pm 0.80$) --- the period of pulsations of RZ~Cas. $P_2 =
0.028354 \pm 0.000646$ $(F_2 = 35.27 \pm 0.80)$ --- almost twice the value of $P_1$.

\section{Conclusions}
We report here photometric observations of RZ~Cas' $\delta$\,Scuti--like pulsations. We stress 
attention to the abrupt increase in amplitude (up to $\sim$0$^{\rm m}$\llap.1) on 12 January, 2005 
that perhaps may be interpreted as a high-mass-transfer-rate event and
inhomogeneity of the accretion stream. Follow-up observations (both
photometric and spectroscopic) of RZ~Cas are strictly desirable
for more detailed study of such events.

\section{Acknowledgments}
Alex Golovin is grateful to the Organizing Committee for  financial
support allowing his participation in the IAU GA. This work was
partially supported by the International Workshop for Astronomy
e.V., enabling Alex Golovin to do part of this project at
Slovakia during 2005 and in the Czech Republic during 2006. It's a
great pleasure for Alex Golovin to express here personal
thankfulness to Jevgeniy Kachalin for his help with preparation of
this manuscript and for the proofreading. We wish to express
thankfulness to D. Mkrtichian for valuable discussions.


\begin{thebibliography}

\bibitem[]{}{Golovin, A. \& Pavlenko, E.} 2006, \textit{JAAVSO}, vol. 34, in press

\bibitem[]{}{Stellingwerf, R.F.} 1978, \textit{ApJ}, 224, 953

\end{thebibliography}
\end{document}